# TRACING OF ERROR IN A TIME SERIES DATA

[1]Koushik Ghosh and [2]Probhas Raychaudhuri

[1]Department of Mathematics
University Institute of Technology
University of Burdwan
Burdwan-713 104
INDIA
Email:koushikg123@yahoo.co.uk

[2]Department of Applied Mathematics
University of Calcutta
92, A.P.C. Road, Kolkata-700 009
INDIA
Email:probhasprc@rediffmail.com

**ABSTRACT:**

A physical (e.g. astrophysical, geophysical, meteorological etc.) data may appear as an output of an experiment or it may contain some sociological, economic or biological information. Whatever be the source of a time series data some amount of noise is always expected to be embedded in it. Analysis of such data in presence of noise may often fail to give accurate information. Although text book data filtering theory is primarily concerned with the presences of random, zero mean errors; but in reality, errors in data are often systematic rather than random. In the present paper we produce different models of systematic error in the time series data. This will certainly help to trace the systematic error present in the data and consequently that can be removed as possible to make the data compatible for further study.

**INTRODUCTION:**

Analysis of data is a very important task since it is the source of information. It provides the validation of theories and models as well as their improvements. Data analysis sometimes can give birth of new theory or model. But in reality the problem is that any kind of time series data either from an experiment or from a dynamical system, or from any economic, sociological or biological aspect usually contains systematic or manual error. Analysis of such data in presence of noise often leads to a wrong interpretation of the data. Analysis of error present in time series data as prescribed in the text books is primarily concerned with the presence of random, zero mean errors. A common treatment in this connection is to consider the error to be Gaussian. But this kind of treatment often fails in reality since the real world problems don't always play by the rules. In reality errors in data are often systematic rather than random. We can believe that there are certain observational processes where the error generated in a certain stage of observation propagates to the next stages and at each stage the effects of preceding errors get enlarged and therefore the errors are arranged in ascending order. We also have certain examples where the process of observation learns from previous experience and as time progresses the system gets more and more efficient to tackle the error. In those cases, any kind of propagation of error is not expected and the errors should be arranged in descending order. If the distribution of error in a time series data is in such well ordered form then we can certainly believe that the error terms are countable and they will not depend upon the corresponding observations. In other words, if the observed data be $\{x_i: i=1, 2, \ldots, n\}$ and the related error terms be $\{\varepsilon_i: i=1, 2, \ldots, n\}$ then these $\varepsilon_i$ depend only on i but not on $x_i$. Tracing of the error in such well ordered form is an interesting study in the field of data analysis. A good tracing of error can give a clear picture of the distribution of error in a time series data. There are also certain cases where these $\varepsilon_i$ depend on $x_i$. For those phenomena these $\varepsilon_i$ are not expected to be well ordered. Tracing of the error terms for those situations can also be performed analytically. In the present analysis we produce some models of distribution of error in order to trace the error in a time series data at different situations.

**THEORY: TRACING OF ERROR:**

We first make an attempt to visualize the situation in which the error terms are arranged in ascending order. Presently our assumption is that the minimum error occurs at the first observation having magnitude a; next it goes on increasing and finally at the last observation the error becomes maximum with magnitude b. We here present four different models of error terms $\{\varepsilon_i: i=1, 2, \ldots, n\}$ embedded in an observed data $\{x_i: i=1, 2, \ldots, n\}$.

**First Model:**

$\varepsilon_1 = a$ and $\varepsilon_i = (b-a) \exp[(n-i)/(1-i)] + a$; $i=2, 3, \ldots, n$  (1)

**Second Model:**

$\varepsilon_i = (b-a) \cos[\pi (n-i)/2(n-1)] + a$; $i=1, 2, 3, \ldots, n$  (2)

**Third Model:**

$\varepsilon_i = b^{(i-1)/(n-1)} a^{(n-i)/(n-1)}$; $i=1, 2, 3, \ldots, n$  (3)

**Fourth Model:**

$\varepsilon_i = a + (b-a)(i-1)/(n-1)$; $i=1, 2, 3, \ldots, n$  (4)

In the first three models we have nonlinear increasing profiles of the error while in the fourth model we produce a linear increasing model.
Now we look for the probable models to visualize the descending profile of error distribution. Here we assume that the maximum error occurs at the first observation having magnitude a; next it moves on decreasing and finally at the last observation the error achieves its minimum value b. We have similar type of models for the present concern:

**Fifth Model:**

$\varepsilon_i = (a-b) \exp[(i-1)/(i-n)] + b$; $i=1, 2, 3, \ldots, (n-1)$ and $\varepsilon_n = b$  (5)

**Sixth Model:**

$$\varepsilon_i = (a-b) \cos[\pi(i-1)/2(n-1)] + b; \quad i=1, 2, 3, \ldots, n \qquad (6)$$

**Seventh Model:**

$$\varepsilon_i = b^{(i-1)/(n-1)} a^{(n-i)/(n-1)}; \quad i=1, 2, 3, \ldots, n \qquad (7)$$

**Eighth Model:**

$$\varepsilon_i = a + (b-a)(i-1)/(n-1); \quad i=1, 2, 3, \ldots, n \qquad (8)$$

As earlier, we have the first three models to depict the nonlinear decreasing profiles and the last one to exhibit a linear decreasing profile.
Although the third and seventh model are identical as well as are the fourth and eighth model but they have altogether different significance.
Finally we draw our attention to the problem where these $\varepsilon_i$ are bound to depend upon the observed data $x_i$. In this context we propose the following model:

**Ninth Model:**

$$\varepsilon_i = \rho \exp(\beta x_i); \quad i=1, 2, 3, \ldots, n \qquad (9)$$

where $\rho$ and $\beta$ are arbitrary positive parameters.
We further assume that the errors at the initial and final observation are known and they are given by $\varepsilon_1 = a$ and $\varepsilon_n = b$. Under that assumption the previous model transforms into

$$\varepsilon_i = a^{(x_i - x_n)/(x_1 - x_n)} b^{(x_1 - x_i)/(x_1 - x_n)}; \quad i=1, 2, 3, \ldots, n \qquad (10)$$

**DISCUSSION: PROPOSED REMEDIAL TREATMENT:**

Next we look for a suitable method of smoothing to reduce the above types of error expected to be present in time series data. In this context we can certainly apply our General Adaptive Rule of Filtering (Ghosh and Raychaudhuri, 2006) for the time series data where the error terms are in well order like in the first eight models. For the observed data $\{x_i\}$; $i=1, 2, 3\ldots, n$ our General Adaptive Rule of Filtering is given by (Ghosh and Raychaudhuri, 2006)

$$y_i = \sum_{k=1}^{i} w_{ki} x_k; \quad i=1, 2, 3, \ldots, n \qquad (11)$$

in which all $w_{ki} \in (0,1)$.
We have restricted ourselves by taking a convex linear combination in (11) and therefore (Ghosh and Raychaudhuri, 2006)

$$\sum_{k=1}^{i} w_{ki} = 1 \qquad (12)$$

We must remember one thing that in order to maintain the positional importance at each stage for a fixed i, $w_{ki}$ must increase steadily with the increase in k so that it can attain its maximum value at k=i. Side by side, these weights are to be taken in such a manner so that the expression for error at each stage can be reduced effectively in the process. We have already established that (Ghosh and Raychaudhuri, 2006) the Simple Exponential Smoothing (Makridakis et al., 1982; Winkler and Makridakis, 1983 and Makridakis and Winkler, 1983) is a favourable example of our method. Conceptually the Simple Exponential Smoothing for a time series data $\{x_i\}$; i=1, 2, 3…, n is given by (Makridakis et al., 1982; Winkler and Makridakis, 1983 and Makridakis and Winkler, 1983)

$y_1 = x_1$ and $y_i = \alpha x_i + (1-\alpha) y_{i-1}$; i= 2, 3…., n where $y_i$ is the smoothed data at the i-th position and α is the parameter of this smoothing. We have established that $\alpha \in (0.5, 1)$ (Ghosh and Raychaudhuri, 2006).

We can easily find that on application of this smoothing in a time series data $\{x_i\}$; i=1, 2, 3…, n where the inherent errors, given by $\{\varepsilon_i\}$; i=1, 2…, n either form a decreasing or an increasing sequence, the total error in the data reduces by

$$\{100/(\sum_{i=1}^{n} \varepsilon_i)\}[\{(1-\alpha)^n - (1-\alpha)\}(\varepsilon_1/\alpha) + \sum_{i=1}^{n-1} (1-\alpha)^{n-i} \varepsilon_{i+1}]\%.$$

For example, if we have the distribution of error like in the fourth model then Simple Exponential Smoothing is able to reduce the total error by $[200 (b-a)(1-\alpha)\{\alpha(n-1) - (1-\alpha) + (1-\alpha)^n\}/\{n(n-1)\alpha^2(a+b)\}]\%$ and that reduction for the eighth model is $[200 (a-b)(1-\alpha)\{\alpha(n-1) - (1-\alpha) + (1-\alpha)^n\}/\{n(n-1)\alpha^2(a+b)\}]\%$.

If the error follows the distribution shown in the ninth model we first arrange the observed data in ascending order and consequently the corresponding error terms will be automatically arranged into ascending order. Next we apply our model of smoothing on that rearranged data and in the newly obtained time series array after smoothing; we again allow a rearrangement to restore the original positions of the data. This follows the desired smoothed time series.

**REFERENCES:**

[1] Ghosh, K. and Raychaudhuri, P., "An Adaptive Approach to Filter a Time Series Data", *Communicated to Management Science* (2006).
[2] Makridakis, S., Anderson, A., Carbone, R., Fildes, R., Hibon, M., Lewandowski, R., Newton, J., Parzen, E. and Winkler, R.L., "The Accuracy of Extrapolation (time series) methods: results of a forecasting competition", *Journal of Forecasting*, **1**, pp. 111-153 (1982).
[3] Makridakis, S. and Winkler, R.L., "Averages of Forecasts: Some Empirical Results", *Management Science*, **29 (9)**, pp. 987-996 (1983).
[4] Winkler, R.L. and Makridakis, S.,"The Combination of Forecasts", *Journal of the Royal Statistical Society, Series A (General)*, **146 (2)**, pp. 150-157 (1983).